\documentclass[aps,prd,showpacs,amsfonts,amssymb]{revtex4}
\usepackage{epsfig}
\def\be{\begin{eqnarray}}
\def\ee{\end{eqnarray}}
\begin{document}
\preprint{FNAL/ ANL-HEP-PR-04-89 } 
\title{Bounding Gauged Skyrmion Masses }

\author{ Yves Brihaye }
 \email{ yves.brihaye@umh.ac.be}
\affiliation{ 
Facult\'e des Sciences, Universit\'e des Mons-Hainaut,
7000 Mons, Belgium}

\author{Christopher T. Hill} 
 \email{hill@fnal.gov }
\affiliation{ 
Fermi National Accelerator Laboratory, P.O. Box 500, Batavia, Illinois 60510, 
USA}

\author{ Cosmas K. Zachos}
 \email{zachos@anl.gov} 
\affiliation{High Energy Physics Division, Argonne National Laboratory,
          Argonne, Illinois 60439-4815, USA}
\date{September 22, 2004} 
\begin{abstract}
Normally, standard (ungauged) skyrmion masses are proportional to the coupling 
of the Skyrme term needed for stability, and so can grow to infinite magnitude 
with increasing coupling. In striking contrast, when skyrmions are 
gauged, their masses are bounded above for 
any Skyrme coupling, and, instead, are of the order of monopole masses, 
$O(v/g)$, so that the coupling of the Skyrme term is not very important.
This boundedness phenomenon and its implications are investigated. 
\end{abstract}
\maketitle

\section{Introduction}

A remarkable feature of theories based upon
$SU(N)_L\times SU(N)_R$ global chiral symmetries is the 
existence of topologically stable field configurations known as
skyrmions. The skyrmion carries a topological charge 
representing the nontrivial homotopy group, $\Pi_3(SU(N))$,
the mapping of the gauge group $SU(N)$ onto a time slice
into the three dimensions of space.
This charge emulates baryon number, and thus skyrmions provide an effective 
model of the baryons of QCD, and their matrix elements
\cite{skyrme,witten,adkins}. 

It is of general interest to consider the skyrmion in the presence
of gauge interactions. There are perturbative gauge interactions in nature
which the skyrmion-as-baryon must experience, i.e., QED, and the electroweak
interactions.  The skyrmion further experiences the 
$\rho$-meson \cite{igarashi}, 
which has an effective description as a gauge field of
isospin. Indeed, gauging chiral Lagrangians promotes the Wess-Zumino term,
which generates the global topological current structure, 
to the Wess-Zumino-Witten term, which not only generates currents
but their anomaly structure as well, and is seen to be
faithful to an underlying theory of quarks and gluons. 
Moreover, if one considers pure Yang-Mills theories
in higher dimensions that undergo compactification to $D=4$, 
one generally finds that the $D=4$ effective description
of KK-modes is a gauged chiral Lagrangian. The skyrmion then
matches higher dimensional
topological objects \cite{decon,hillramond,hill}. 
There have also been interesting applications 
of gauged skyrmions in the context of technibaryon decay \cite{rubakov}.

In the present paper, we examine the impact
of the diagonal gauging (i.e., the promotion of the diagonal
subgroup $SU(2)$ (isospin) of the chiral group
$SU(2)_L\times SU(2)_R$ to a gauge group).  
In particular, we observe novel behavior
for the masses of gauge skyrmions which significantly
departs from the global case. 

The conventional (ungauged) skyrmion is a
solution to the chiral model equations of motion
 \cite{skyrme,zahed}
supplemented with a ``Skyrme term.'' The Skyrme term is required to
stabilize the core structure, but one finds a 
sensitivity to the strength of this term in the mass: the mass of
the skyrmion is essentially proportional to the 
square-root of the coupling coefficient of the Skyrme term. This 
term has to be input, by hand, or to be somehow justified as emerging 
from a long-distance effective
Lagrangian description of a more complicated system, 
e.g., from some shorter distance scale physics in QCD.  

The skyrmion solutions of particular gauged chiral models, however,  
exhibit unexpected behavior, even though tantalizing hints of it
could be gleaned from pioneering numerical studies of gauged skyrmions 
\cite{arthur,brihayeF,brihayeG,brihayeH}. Specifically, the mass of the 
skyrmion increases monotonically from zero with the Skyrme coupling, 
but does not go to infinity as it would for ungauged skyrmions. 
Instead, the mass stabilizes to an upper bound, whose scale, $O(v/g)$, is 
``monopolic", i.e., it is set by the gauge coupling and the characteristic 
spontaneous symmetry breaking scale. This limit conforms to 
the masses of magnetic monopoles, 
which likewise do not vary much 
above the minimal BPS values \cite{kirkman}.
 
For simplicity, we focus on plain Skyrme-Wu-Yang spherically symmetric 
hedgehogs. We discuss the simplest system 
\cite{rubakov,hill},  $SU(2)_L \times  SU(2)_R$ with gauged 
diagonal $SU(2)_V$,   
\be 
4 \pi E(g,v,\kappa)  \equiv
\frac{1}{2}  \int \!\! d^3x {\rm Tr} F_{ij} F_{ij} 
+\frac{v^2}{2} \int \!\! d^3x \Bigl (  
{\rm Tr}  [ D_j, U^\dag ] [ D_j , U] +\kappa^2 {\rm Tr}  \left (  
[ [ D_j, U^\dag ],  [ D_i , U] ] \right )^2  \Bigr ) ~, \label{startup} 
\ee

The spherical hedgehog Skyrme-Wu-Yang Ansatz\cite{skyrme,wuYang}  
(i.e. of unit winding/``baryon" number) is 
\be
A_i= \frac{a(r)-1}{gr}~ \epsilon _{ijk}\frac{\tau^j}{2} \hat{x}^k,  
\qquad \qquad U=\exp \Bigl ( i f(r) ~\hat{\mathbf{x}} \cdot \tau \Bigr )
= \cos f(r) 
+ i\hat{\mathbf{x}} \cdot \tau ~\sin f(r).
\ee
(The exponent $f(r) ~\hat{\mathbf{x}} \cdot \tau \sim \int dx^4 ~ A_4$, 
amounts to the deconstruction Wilson line/link  \cite{hill}.) 

This two-scale problem yields an energy $E(g,v,\kappa)$ which 
has a lower, topological bound  \cite{brihayeG}. Moreover, it is 
manifestly {\em monotonic} in the Skyrme coupling strength $\kappa$, 
because the $\partial / \partial \kappa$ derivative is positive 
semidefinite, while all implicit dependence of the fields on $\kappa$ 
vanishes on-shell (by use of the eqns of motion), and is thus irrelevant, 
as in the case of the monopole mass varying as a function of the
Higgs mass \cite{kirkman}. 

To familiarize the reader with the bounding arguments, we first summarize the 
standard results on the simplest, $B=1$, ungauged skyrmion \cite{skyrme}
in Section II. We then define and examine the simplest, $B=1$, gauged skyrmion; 
we review lower bounds, 
$E_{topological}$  \cite{arthur,brihayeF,brihayeG,brihayeH};  
and, finally, we  
derive upper bounds for its mass, on the basis of numerical investigation
for asymptotically large couplings, in Section III. Asymptotically, we 
find this actual (bounding) mass of the $B=1$ simple skyrmion to be merely 
$2.06~ E_{topological}$. 
Our results are qualitatively unchanged upon further introduction of a 
pion mass \cite{brihayeH}---even though, as expected, the corresponding 
upper bound increases with the mass of the pion.  In the last Section, 
IV, we conclude with discussion and interpretation of the phenomenon.

\section{Review of Lower bounds of Ungauged Skyrmions} 

In the limit of decoupling of the gauge fields, $g=0$, or equivalently,
$a=1$, (\ref{startup}) yields the standard single skyrmion  reviewed here. 
It is evident from scaling, below, that all activity occurs at scales of 
$r= O(\kappa)$, so that $E=O( \kappa  v^2)$: 
\be
E(0,v,\kappa)&=& \frac{v^2}{2} \int_0^\infty dr \left ( r^2  
f'^2 + 2 \sin ^2  f
+ \kappa^2 \frac {\sin ^4  f}{r^2}
+ 2 \kappa^2 ~f'^2 ~ {\sin ^2  f} \right )\\
&=& \frac{v^2}{2} \int_0^\infty dr~  \left (  
(rf' -\kappa \frac {\sin ^2  f}{r})^2 
+ 2 \sin ^2  f ~ (1-\kappa f' )^2 +
6 \kappa f' \sin ^2  f\right ) . \nonumber 
\ee

The first two terms in the integrand are positive semi-definite. 
The last one is a total divergence, $12 \kappa \pi^2 r^2 \times$ the 
topological (Chern-Simons) baryon density of the conventional skyrmion
\cite{witten}, 
\be
\frac{3\kappa}{2} \partial_r (2f - \sin 2f)= \frac{\kappa r^2 }{2} 
\epsilon^{ijk} {\rm Tr} ( U^\dag \partial_i U ~  
U^\dagger \partial_j U ~  
U^\dagger \partial_k U )~.  
\ee
Thus, its contribution to the energy is 
\be
E_{topological}=   \frac{3v^2 \kappa}{4} \Big ( 
2 f(\infty) -2f(0) - \sin 2f(\infty) +
\sin 2f(0) \Big ) = \frac {3 \pi}{  2}  ~\kappa v^2 . 
\ee
Note that one could choose either sign in completing the above squares.

This is a Bogomol'ny {\em topological lower bound}. However, it cannot be 
saturated, as it does in the BPS monopole case (There are no self-dual 
chiral fields \cite{zahed}). Saturation would require both squares to vanish, 
which is impossible: ~ both $f'=1/\kappa$, and  $\kappa \sin f = r$.

This lower bound melts away for vanishing Skyrme term $\kappa=0$, and
blows up for $\kappa \rightarrow \infty$. 
Thus, there can be no $\kappa \rightarrow \infty$ upper bound for 
the ungauged skyrmion. 

\section{Gauged Skyrmions and their upper and lower bounds} 
By contrast, when gauge fields 
are introduced, $g\neq 0$,  $\kappa$-dependence drops out of the 
lower bound as $\kappa \rightarrow \infty$; moreover, as we show below, 
{\em there is an  upper bound for the gauged skyrmion} above, remarkably close 
to the highest lower bound. 

The full equation (\ref{startup}) then for the $B=1$ Ansatz, 
with $r$ and $\kappa$ rescaled in units of $gv$, amounts to 
\be
E(g,v,\kappa)=   
\frac{v}{g} \int_0^\infty \!\! dr \left (  
4 a'^2 + \frac{2 (a^2-1)^2}{r^2} +
\frac{r^2 f'^2}{2} + a^2 {\sin ^2  f}
+ \kappa^2 a^4 \frac {\sin ^4  f}{2r^2}
+ \kappa^2 ~a^2 f'^2 ~{\sin ^2  f}\right ),  \label{mother}
\ee
with boundary conditions \quad  $a(0)= 1, \quad a(\infty)= 0; 
\quad f(0)= 0, \quad f(\infty)= \pi$.

The topological lower bound \cite{brihayeG} is 
\be 
E_{topological} > \frac{2 \pi v}{\sqrt{g^2+(\frac{4}{ 3 \kappa v})^2}} ~,
\ee
which yields, in the limit $\kappa \rightarrow \infty$, 
a lower bound of $O(v/g)$, i.e., of the order of the BPS monopole mass. 
Once gauging is switched on, $1/g^2$ and $\kappa^2$ behave
analogously to resistors in parallel: as $\kappa$ blows up, it 
becomes irrelevant, leaving the scale to be set by $g$. 

The gauged skyrmion has been well studied numerically 
in refs \cite{arthur,brihayeF,brihayeG,brihayeH}, which detail a 
remarkable branch structure of solutions. For increasing Skyrme 
coupling $\kappa$, the energy of the simple ($B=1$) skyrmion is seen to 
increase, starting from 0. (Specifically, Fig 1 of ref \cite{brihayeG}, 
for increasing $\kappa g v$, the gauged skyrmion energy $E$ curves 
over, in contrast to that of the ungauged skyrmion on the same graph.) 
In fact, we show that it flattens out asymptotically, yielding an upper bound,
 as is the case for the 'tHooft-Polyakov monopole as a function of 
the Higgs mass \cite{kirkman}. This bound is very close to the highest lower 
bound.
 
The full lower bound for the gauged case results from rewriting
the above energy as 
\be
E(g,v,\kappa)&=& \frac{v}{g} \int_0^\infty \!\! dr \Biggl (  \Bigl 
(4a'^2+\frac{(3\kappa/4)^2}{1+(3\kappa/4)^2} \frac{a^2 \sin^2  (2f)} {4}\Bigr )
+  \frac{(3\kappa/4)^2}{1+(3\kappa/4)^2} a^2 \sin^4 f 
+ a^2 \sin^2 f \Bigl (\frac{1}{1+(3\kappa/4)^2}+ \kappa^2 ~f'^2\Bigr )
\nonumber \\ 
&+& \frac{1}{2}  \Bigl (\frac{r^2 f'^2} {1+(3\kappa/4)^2}+ 
\kappa^2 a^4 \frac {\sin ^4  f}{r^2} \Bigr    ) 
+ \frac{1}{2}  \Bigl (  \frac{(3\kappa/4)^2}{1+(3\kappa/4)^2}
r^2 f'^2 +\frac{4 (a^2-1)^2}{r^2} \Bigr ) \Biggr )\nonumber\\ 
&=& \frac{v}{g} \int_0^\infty \!\! dr \Biggl  (  
\Bigl  (2a'+ \frac{(3\kappa/4)}{\sqrt{1+(3\kappa/4)^2}} 
\frac{a \sin  (2f)} {2}\Bigr  )^2     
+  \frac{(3\kappa/4)^2}{1+(3\kappa/4)^2} ~a^2 \sin^4 f \nonumber \\
&+& a^2 \sin^2  f \Bigl   (\frac{1}{\sqrt{1+(3\kappa/4)^2}}   -\kappa ~f'
\Bigr  )^2 
+ \frac{1}{2} \Bigl (\frac{r f'} {\sqrt{1+(3\kappa/4)^2}}- \kappa a^2 
\frac {\sin ^2  f}{r} \Bigr) ^2 \nonumber\\ 
&+&
\frac{1}{2} \Bigl   (r f' \frac{(3\kappa/4)} {\sqrt{ 1+(3\kappa/4)^2}}
+ \frac{2 (a^2-1)}{r}\Bigr   )^2 \Biggr ) ~+ ~
\frac{3\kappa v}{4g\sqrt{1+(3\kappa/4)^2}}
\int_0^\infty \!\! dr ~ \partial_r (  2f - a^2 \sin 2f ) \nonumber\\ 
&>& \frac{2\pi v}{g\sqrt{1+(4/3\kappa)^2}}~~,   \label{lowerbound}
\ee
since each term but the last (surface term) is positive semi-definite.
As before, all these terms cannot be nullified simultaneously,
and thus the topological bound is not saturated, except in the degenerate case 
$\kappa=0$, cf.\ \cite{brihayeG}. The highest value 
for this lower bound, $2 \pi v/g$, holds for $\kappa \rightarrow \infty$; it 
will be seen that the actual energy is roughly twice this, in that limit.

The Euler-Lagrange equations are
\be
a''+ \frac{a(1-a^2)}{r^2} -\frac{a}{4} \sin^2 f     -\frac{\kappa^2  a}{4} 
f'^2 \sin^2 f     -\frac{\kappa^2 a^3  }{4r^2} \sin^4 f =0,  \label{eqvec}
\ee
\be
(r^2 + 2 \kappa^2 a^2   \sin^2 f ) f'' + 2rf' 
 +4\kappa^2   aa'  f'\sin^2 f +\kappa^2 a^2  f'^2  \sin (2f)   
 -a^2 \sin (2f)  -\frac{2 \kappa^2  a^4 }{r^2} \sin^3 f  \cos f 
  =0,  \label{eqscal}
\ee
with BCs: 
\be
a(0)= 1,  \qquad 
a(\infty)= 0; \qquad \qquad   
f(0)= 0, \qquad 
f(\infty)= \pi. 
\ee 

Study of actual numerical solutions for increasing $\kappa^2$, 
(cf., e.g., branch B of ref \cite{brihayeG}), 
reveals that the scales of the two variables $a$ and $f$ tend to resolve 
as in Born-Oppenheimer problems, and the monopole in the large Higgs mass 
limit \cite{kirkman}. Specifically, $a$ asymptotes 
faster than $f$, which continues to evolve slowly after $a$ has 
decayed to 0 by some value $r=R$. Numerically, $R \sim 1.34$,
very weakly dependent on $\kappa$, for large $\kappa^2 \sim 1000$
(cf.\ Fig.\ {\ref{fig1}).
\begin{figure}[htb]
\centering
\epsfysize=18cm   
\mbox{\epsffile{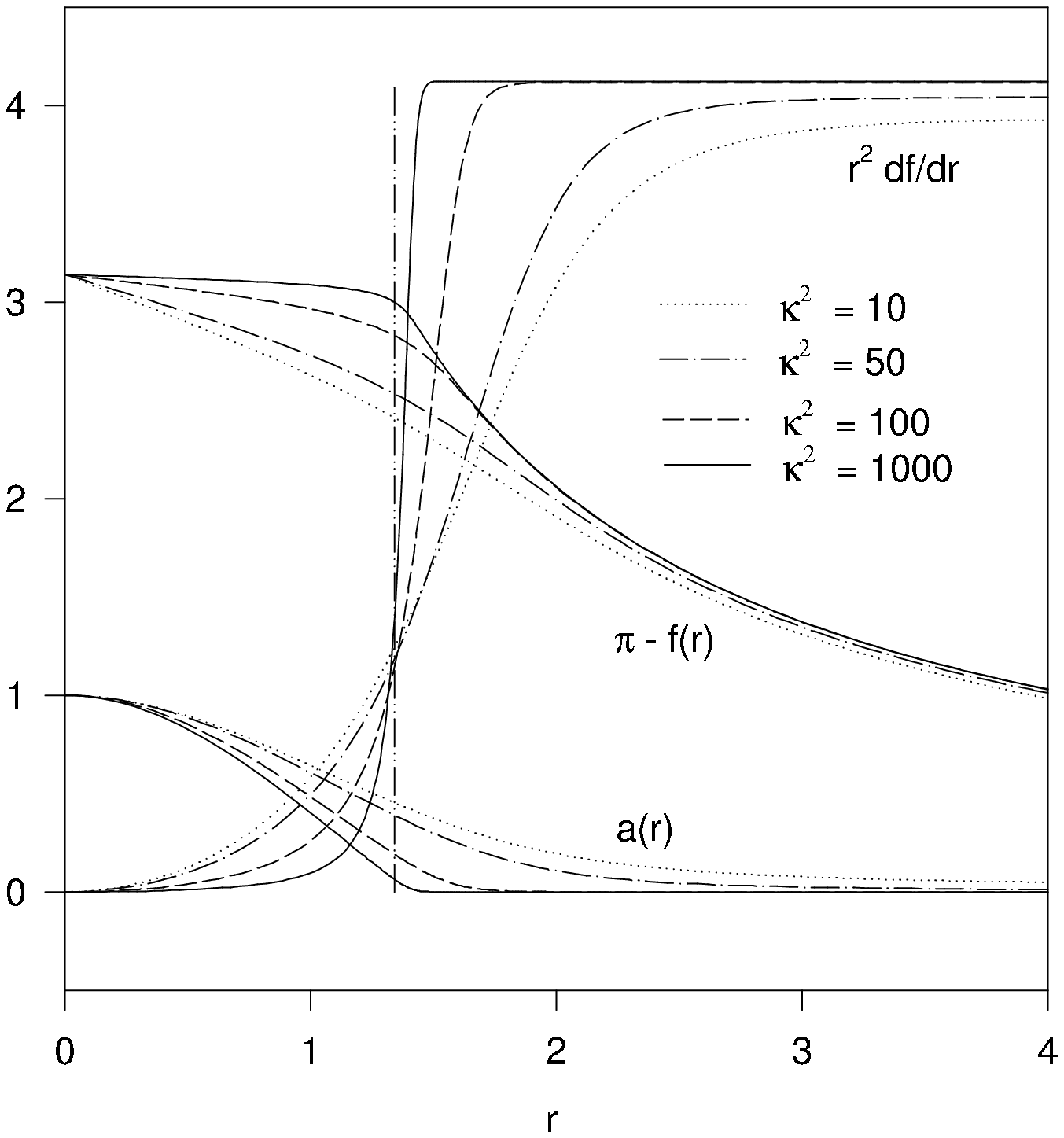}}
\caption{\label{fig1} 
Numerical study of $a$;  $\pi -f(r)$; and 
$r^2 f'$, the corresponding argument of eqn (\ref{tail}),
for increasing values of $\kappa^2$. It is evident that
the two variables tend to decouple, for sufficiently large $\kappa^2$: 
the scales of $a$ and $f$ separate at $R \sim 1.34$,
demarcated by the vertical dash-double-dot line. 
Note the abrupt transition of $f'$ from null to $\pi R /r^2$ behavior.}
\end{figure}

Specifically, for the proximate interval,  $[0,R ]$,
\be
f(r)\sim 0,  
\ee 
so that eqn (\ref{eqscal}) collapses, while (\ref{eqvec}) effectively reduces 
to 
\be
a''+ \frac{a(1-a^2)}{r^2} \sim 0,   \label{eqlim}
\ee
the celebrated Wu-Yang equation\cite{wuYang} for pure Yang-Mills 
theory. (Note, however, that this equation, by itself, is 
scale invariant: the actual scale $R$ is set through interaction 
with $f$. The range of $a$ would spread out, left to itself, but the Skyrme
term disfavors overlap of $a$ with $f$. In the limit, it forces $a$ to 
attenuate inside the core, before $f$ builds up at $R$.)
Hence $a(r)$ is an attenuating function which  reaches $a(R)\sim 0$. 

For the distant interval $[R, \infty )$, ~~~$a\sim 0$, so that 
(\ref{eqvec}) collapses, while  (\ref{eqscal}) 
reduces to  
\be
\partial_r ( r^2 f')\sim 0 ,   \label{tail}
\ee
and hence $f\sim \pi (1- \frac{R}{r} )$.  

Thus, for $\kappa\rightarrow \infty$, dependence on $\kappa$ dies out. For 
solutions (on-shell),
the coefficient of $\kappa^2$ in the energy collapses, 
\be
\frac{dE}{d(\kappa^2)}=\frac{\partial E}{\partial (\kappa^ 2)}= 
\int_0^{\infty}  dr ~\left ( a^2 f'^2 \sin^2 f 
+ \frac{a^4 \sin^4 f}{2 r^2 } \right ) \sim 1.62 ~ \kappa^{-5/2}
\rightarrow 0.  \label{leadcoeff}
\ee

In the most important region, the neighborhood of $R$, activity is 
apparently dominated by the scale $r\sim \sqrt {\kappa}$; this is 
quite unlike the characteristic scale of the ungauged  
skyrmion activity, $r\sim \kappa$. 

Since $E$ is monotonic in $\kappa$, an upper 
bound results for (\ref{mother}) in this limit,
\be
E(g,v,\kappa)\leq  E(g,v,\infty) \sim 12.95~ \frac{v}{g} ~ .  \label{bound}
\ee
This upper bound is merely $\frac{g}{v} E(g,v,\infty)\sim 2.06 \times 2\pi$, 
where $2\pi$ is the above-mentioned highest lower bound. 
In effect, the mass of the gauged skyrmion varies from 0 
to 12.95 $v/g$, as the Skyrme
term ranges from zero to infinite strength. Near zero, the Skyrme coupling 
$\kappa$ sets the mass scale, but for large couplings the scale is 
set by the ``monopole mass" scale $v/g$.

In more numerical detail, for $\frac{g}{v} E(g,v,\infty)\sim  12.95$, 
the subleading behavior is  
\be
E(g,v,\kappa)= E(g,v,\infty) - 6.68 ~\frac{v}{g}~\kappa^{-1/2} +O(\kappa^{-1}).
\ee

In Fig.\ 2, beyond $E$,  $E_{sk}$ is also plotted. It represents the 
``Skyrme term", i.e., the last two terms in eqn (\ref{mother}). The decay of 
$E_{sk}$ goes like $E_{sk}\sim 1.62 \kappa^{-1/2}$, as indicated, so 
this component is subdominant to the contributions of the Wu-Yang and 
the conventional chiral terms, the leading four terms in eqn (\ref{mother}).
Since 
\be
E_{sk}= \kappa^2 \frac{dE}{d(\kappa^2)}=
-\frac{1}{4} \kappa^{-1/2}  \frac{dE}{d(\kappa^{-1/2})},
\ee
this is seen to be numerically consistent with the above expansion in 
$\kappa^{-1/2}$ around $\kappa^{-1/2}=0$. 
  
\begin{figure}[htb]
\centering
\epsfysize=18cm   
\mbox{\epsffile{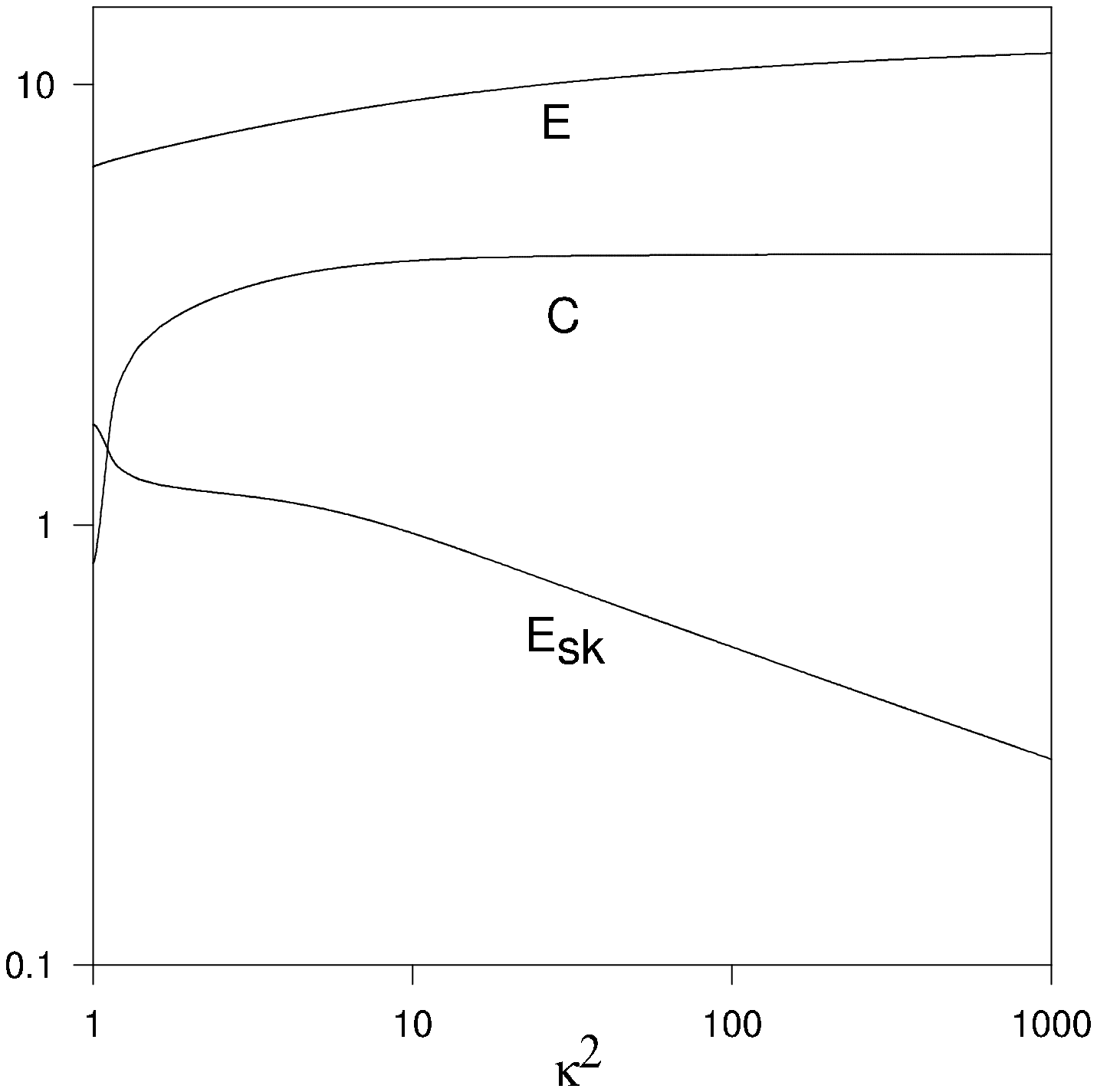}}
\caption{\label{fig2} 
Numerical study of the energy $E$, in units of 
$v/g$, for increasing $\kappa^2$. The upper bound is at 
$E(g,v,\infty)\sim  12.95$.  
Also plotted is  $E_{sk}$, representing the last two terms in 
eqn (\ref{mother}),  the Skyrme term. 
Further plotted is the coefficient $C$ of the leading tail of 
$f=\pi (1 -C/r + O(1/r^2) )$ for large $r$, tending to $\pi R$ 
in the limit $\kappa \rightarrow \infty$. It may be instructive to note the 
contrast to the $\pi -2.16/r^2$ asymptoting of the conventional 
ungauged skyrmion.}
 \end{figure}

The boundary condition for $f(r)$ above may be effectively regarded as a 
``unit baryon charge constraint" \cite{brihayeH}; to enforce it more naturally,
it is customary to add a pion mass term, $\lambda r^2 ( 1- \cos f) $ to the 
integrand of (\ref{mother}), arising out of a term 
${\rm Tr} (U + U^\dagger )-4$ in the chiral Lagrangian, where 
$\lambda= 2 (m_\pi / gv)^2$. However,
addition of such a term does not alter the qualitative conclusions 
above. 

For example, for $\lambda=1$, the upper bound is only somewhat higher,
\be
\frac{g}{v} E(g,v,\kappa)= 17.5 - 6.0~~\kappa^{-1/2} +O(\kappa^{-1}),
\ee
as expected: since this mass term is positive-semidefinite, the skyrmion 
mass is a monotonically increasing function of $\lambda$.

\section{Discussion} 
Our present analysis probed the effects on
skyrmionic masses of the presence of the
diagonal gauge group. The situation is reminiscent
of the 't Hooft-Polyakov monopole,  for which
a similar bound exists owing to  gauge
fields \cite{kirkman}. 
While the monopole is a tangle of gauge, Goldstone, and Higgs fields; 
as the mass of the Higgs field is
taken to infinity with fixed VEV, the Higgs serves only to 
enforce boundary conditions. The monopole in this limit ends up made 
purely of gauge 
fields (a Higgsless monopole, as in ref \cite{vinciarelli}), with a mass 
of $O(M_W/\alpha)$.  Analogously, the gauged skyrmion in 
(\ref{mother}) consists of gauge fields and Higgs field skyrmions.
But, in the interaction with very heavy would-be skyrmions, the last two 
terms in that system, the ``Skyrme term" (viz.,  $E_{sk}$ of Fig 2), 
become decreasingly relevant in the energy. Thus, 
what would have been the infinitely massive
skyrmion largely enforces boundary conditions
at $R$. The leading two terms in the energy (the gauge, or Wu-Yang, part)
scale as $1/s$ with $r\rightarrow sr$, and, left to themselves, favor 
a spread-out integrand to maximize $s$. 
The next two terms (the chiral action terms) scale as $s$, and favor 
core-shrinking, but the last two terms (the Skyrme terms) 
oppose this, and stabilize the core to $\sim R$, thereby constraining the 
gauge field within this range.  The mass of the
skyrmion ends up of the order 
characteristic of monopole configurations, $O(v/g)$, 
superficially oblivious of the Skyrme coupling.

We have thus found the effects of gauging to be
significant in the limit of large Skyrme term coefficient.
We note, however, that one could \cite{hill} (and should \cite{future}) 
include the effects of additional  operators that explicitly
involve the Yang-Mills field strength, and are of the
same dimension as the gauged Skyrme term utilized here, such as 
${\rm Tr} F_{ij}[ D^j, U^\dag ] [ D^j , U] + {\rm h.c}$.
There are potentially interesting effects of these, to be
considered elsewhere \cite{future}. 

\section*{Acknowledgments}
We wish to record our obligation to D.~H.~Tchrakian and to D.~B.~Fairlie for 
helpful conversations. 
This work is supported in part by the Belgian FNRS; 
the US Department of Energy, High Energy Physics Division, 
under Grant DE-AC02-76CHO3000; and Contract W-31-109-ENG-38.

\end{document}